\documentclass[epj,referee]{svjour}
\usepackage{graphics}
\usepackage{epsfig}

\begin{document}

\title{Convective and Absolute Instabilities in the Subcritical
Ginzburg-Landau Equation}

\author{Pere Colet \and Daniel Walgraef\thanks{Permanent address: 
Centre for Non-Linear Phenomena and Complex Systems, Universit\'e Libre de
Bruxelles, Campus Plaine, Blv. du Triomphe B.P 231, B-1050 Brussels, Belgium} 
\and Maxi San Miguel}

\institute{Instituto Mediterr\'aneo de Estudios Avanzados, 
IMEDEA\thanks{Electronic address: {\tt http://www.imedea.uib.es/PhysDept/}}
(CSIC-UIB),
 Campus Universitat Illes Balears, E-07071 Palma de Mallorca, 
Spain.}

\date{11 December 1998}

\abstract{
We study the nature of the instability of the homogeneous steady states of
the subcritical Ginzburg-Landau equation in the presence of group
velocity. The shift of the absolute instability threshold of the trivial
steady state, induced by the destabilizing cubic nonlinearities, is
confirmed by the numerical analysis of the evolution of its perturbations.
It is also shown that the dynamics of these perturbations is such that
finite size effects may suppress the transition from convective to
absolute instability. Finally, we analyze the instability of the
subcritical middle branch of steady states, and show, analytically and
numerically, that this branch may be convectively unstable for
sufficiently high values of the group velocity.
}

\maketitle

\section{Introduction}

Several physico-chemical systems driven out of equilibrium present
stationnary instabilities of the Turing type, or oscillatory instabilities
corresponding to Hopf bifurcations. Such instabilities lead to the
formation of various kinds of spatio-temporal patterns\cite{crosshoh}.
Well known examples are: Rayleigh-B\'enard  instabilities in Newtonian
fluids, binary mixtures, or viscoelastic solutions\cite{binary,larson},
electrohydrodynamic instabilities in nematic liquid crystals \cite{ehd},
Turing instabilities in nonlinear chemical systems \cite{shoka},
convective instabilities in Taylor-Couette devices \cite{taylorcouette},
etc. Close to such instabilities, the dynamics of the system may usually
be reduced to amplitude equations of the Ginzburg-Landau type, which
describe the evolution of the patterns that may appear beyond the
bifurcation point \cite{daniel}.

According to the system under consideration, and to the nature of the
instability, these Ginzburg-Landau equations may contain mean flow terms 
induced by group velocities. In this case, pattern formation crucially
depends on the convective or absolute nature of the instability. Let us
recall that, when the reference state is convectively unstable, localized
perturbations are driven by the mean flow in such a way that they grow in
the moving reference frame, but decay at any fixed location. On the
contrary, in the absolute instability regime, localized  perturbations
grow at any fixed location \cite{huerre}. The behavior of the system is
thus qualitatively very different in both regimes. In the convectively
unstable regime, a deterministic system cannot develop the expected
patterns, except in particular experimental set-ups, while in a stochastic
system, noise is spatially amplified and gives rise to noise-sustained
structures \cite{deissler,ahlers,marco}. On the contrary, in the
absolutely unstable regime, patterns are intrinsically sustained by the
deterministic dynamics, which provides the relevant selection and
stability criteria \cite{mueller,buechel}. Hence, the concepts of
convective and absolute instability are essential to understand the
behavior of nonlinear wave patterns and their stability
\cite{deissler,mutveit}.

The nature of the instability of the trivial steady state has been
studied, either numerically, analytically and experimentally: In the case
of supercritical bifurcations, linear criteria are appropriate to 
determine the absolute instability threshold, and to analyze the
transition from convective to absolute instability
\cite{deissler,ahlers,marco,deisslerbrand,helmutda,steinberg,luecke,marc}. 
However, in the case of subcritical bifurcations, the nonlinearities are 
destabilizing, which leads to the failure of linear instability criteria.
In a qualitative analysis based on the potential character of the real
subcritical Ginzburg-Landau equation, Chomaz \cite{chomaz} argued that the
transition between convective and absolute instability of the trivial
steady state should occur at the point where a front between the rest
state and the nontrivial steady state is stationary in a frame moving with
the group velocity. This defines the nonlinear convective-absolute
instability threshold, above which nonlinear global modes are
intrinsically sustained by the dynamics, as discussed by Couairon and
Chomaz \cite{couairon}. This argument relies on the existence of a unique
front between the basic and the bifurcating states, as it is the case in
the subcritical domain where both basic and bifurcating states are
linearly stable. In the supercritical domain, where the basic state is
linearly unstable, or in the complex Ginzburg-Landau equation, this front
is not unique any more, and, as commented by van Hecke et al.
\cite{vanhecke}, one has to know which nonlinear front solution is
selected, to determine the nonlinear stability properties of the basic
state.

Within this context, our aim in this paper is to contribute to the study
of this problem addressing some aspects of it that so far have not been
considered. Additionally we study stochastic effects. A first aspect
concerns finite size effects and their influence on the transition from
convective to absolute instability for the trivial steady state. Indeed,
our numerical analysis of the evolution of perturbations of this state
show that it consists in two stages. The first one is devoted to the
building of a front between this state and the bifurcating one. It is
during the second stage that this front moves outwards or inwards
according to the convective or absolute nature of the instability. We will
show that, although the absolute instability threshold may effectively
be shifted, due to nonlinear effects, in agreement with \cite{chomaz}, the
first step of the evolution is sensitive to the size of the system, and
this may affect the practical determination of the absolute instability
threshold, and even suppress it. 

A second aspect is the effect of the group velocity on the unstable
subcritical branch. In the subcritical domain, there is a middle branch
of steady states, between the trivial and the bifurcating ones. In fact,
the nature of the instability of this branch in the presence of a group
velocity has not been considered so far. In the absence of group
velocity, this branch is absolutely unstable. However, the nature of the
instability may be modified in systems with group velocity or mean flow
effects. We will effectively show, analytically and numerically, that
this unstable subcritical branch may be convectively unstable, totally or
partly, according to the mean flow intensity. Effectively, in
deterministic systems, unstable states on this branch do not necessarily
decay in the presence of group velocity, while they may remain long lived
in stochastic systems. This fact may be of practical importance, since it
provides an alternative way to stabilize the subcritical middle branch,
which is qualitatively different from the one proposed by Thual and Fauve
\cite{thualfauve}. It could, furthermore, provide the last building block
needed for the understanding of pattern formation in binary fluid
convection, as suggested in \cite{crosshoh}.

In section \ref{subcriticalGL}, we recall the dynamical model. In section 
\ref{trivialstate} we discuss the nature of the instability of the 
trivial steady state, and present the results of a numerical analysis of
the problem. In section \ref{bifurcatingstate}, we show, analytically and
numerically, that the subcritical middle branch of steady states may be
convectively unstable, and may thus be stabilized by mean flow effects in
deterministic systems. Finally, conclusions are drawn in section
\ref{conclusions}.

\section{The Subcritical Scalar Ginzburg-Landau Equation}
\label{subcriticalGL}

For the sake of simplicity, we will consider, in this paper, systems
described by a scalar order parameterlike variable, and where the
dynamics is given by the real fifth-order Ginzburg-Landau equation, which
may be written, in one-dimensional geometries, as \cite{crosshoh,wimhoh}:
\begin{equation}\label{Rcglsub} 
\partial_t A + c\partial_x A = \epsilon A 
+ \partial^2_xA + v A ^3 - A^5 +\sqrt{\xi} \chi (x,t), 
\end{equation}
For future reference we have added to the equation a stochastic term $\chi
(x,t)$. This models a Gaussian white noise of zero mean and variance
given by $\left< \chi (x,t) \chi (x',t') \right> = 2 \delta (x-x')\delta
(t-t')$. In the remainder of this section we consider the deterministic
situation with $\xi=0$.

Bifurcating uniform steady states $A(x,t)=R$ of this equation are well
known:
\begin{equation}\label{cglsubampl}
R^2_{\pm} = {1\over 2}(v \,\pm \,\sqrt { v^2 + 4\epsilon}) \,.
\end{equation}

\begin{figure}
\hspace{-2cm}
\epsfig{figure=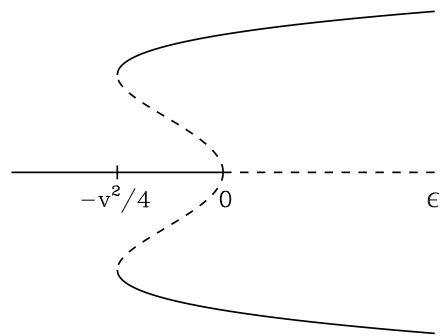,width=10cm}
\vspace{-1.5cm}
\caption{Bifurcation diagram for the real subcritical scalar Ginzburg-Landau 
equation.\label{bifurcation}}
\end{figure}

The linear evolution of the perturbations $\rho_{\pm} = A - R_{\pm}$ around 
these states is then given by:
\begin{equation}\label{linampl}
\partial_t \rho_{\pm} + c\partial_x \rho_{\pm}= \mp 2R^2_{\pm}\sqrt { v^2 + 
4\epsilon}\,\rho_{\pm} 
+ \partial^2_x\rho_{\pm} \,. 
\end{equation}
Hence, in the absence of group velocity, the upper branch $R_+$ exists and
is stable for $-{ v^2 \over 4}<\epsilon$, while the middle branch $R_-$
exists and is unstable for $-{ v^2 \over 4}<\epsilon<0$ (cf. fig.
\ref{bifurcation}). 

This picture is, of course, known to change in the presence of a finite
group velocity $c$. Let us first recall the linear and nonlinear criteria
for convective and absolute instability for the trivial steady state
$A=0$.

\section{Linear and Nonlinear Instability of the Trivial Steady State}
\label{trivialstate}

\subsection{Analytical Results}

The linear evolution around the trivial steady state $A=0$ is given by 
\begin{equation}
\partial_t \rho_0 + c\partial_x \rho_0 = \epsilon \,\rho_0 
+ \partial^2_x\rho_0
\end{equation}
and the corresponding dispersion relation is 
\begin{equation}
\omega = \epsilon - c\kappa + \kappa^2
\end{equation}
with $\kappa = k' + i k''$. The usual linear instability criterion 
\cite{huerre} 
\begin{equation}
\Re {d\omega\over d\kappa} = \Im {d\omega\over d\kappa} = 0
\end{equation}
and $\Re(\omega(\kappa))=0$ gives that the trivial steady state is 
convectively unstable for $0 < \epsilon < c^2/4$, and absolutely unstable for
$\epsilon > c^2/4$.

However, since the nonlinearities of the dynamics are destabilizing, the
linear terms may possibly not govern the growth of perturbations of the
steady state. Hence, a reliable stability analysis has to include
nonlinear terms. As discussed by Chomaz and Couairon
\cite{chomaz,couairon}, the nonlinear stability analysis of the trivial
steady state relies on its response to perturbations of finite extent and
amplitude. Hence, in the case eq. (\ref{Rcglsub}), without group velocity
($c=0$), it is sufficient to consider a front solution joining the $0$
state at $x \to - \infty$ to the $R_+$ state at $x \to + \infty$. 

\begin{figure}
\epsfig{figure=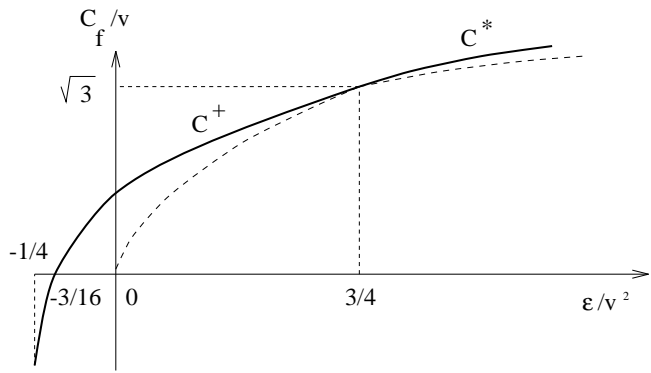,width=8.5cm}
\caption{Representation of the linear and nonlinear front velocities 
($c^*$ and $c^{\dagger}$) in the ($c_f/v$, $\epsilon/v^2$) plane. The
solid line represents the selected front velocity. The dashed lines
represent $c^*$ for $\epsilon < 3v^2/4$ and $c^{\dagger}$ for
$\epsilon > 3v^2/4 $
\label{frontvelocity}}
\end{figure}

In the case of the dynamics given by Eq. (\ref{Rcglsub}), this velocity,
$c_f$ may be calculated exactly \cite{wim}, and is found to be (cf. fig.
\ref{frontvelocity})
\begin{eqnarray}
c_f &=& c^{\dagger} = {1\over \sqrt 3}(-v+2\sqrt{v^2 + 4\epsilon}) \, 
({\rm for}\, -{v^2\over 4} <\epsilon < {3v^2\over 4})\nonumber \\
c_f &=& c^* = 2\sqrt{\epsilon} \, ({\rm for}\,
 {3v^2\over 4} <\epsilon )
\label{cfront}
\end{eqnarray}
Note that $c^*$ is the linear marginal velocity.

If the front velocity is negative, which is the case for $\epsilon <
-3v^2/16$, an isolated droplet of the $R_+$ state embedded into the $0$
state shrinks, and the $0$ state is stable. On the contrary, if $c_f$ is
positive, which is the case for $\epsilon > -3v^2/16$, $R_+$ droplets
grow, and the $0$ state is nonlinearly unstable. The value $\epsilon =
-3v^2/16$ corresponds to the Maxwell construction of phase transitions in
which the trivial and upper branch have equal stability.

When $c \ne 0$ and $v=1$, Chomaz \cite{chomaz} showed that, in the
unstable domain ($\epsilon > -3v^2/16$), the instability is nonlinearly
convective (NLC) when $ c_f < c$, since, in this case, although
expanding, a $R_+$ droplet is finally advected out of the system. On the
contrary, when $ c_f > c$, the instability is absolute (NLA), since, in
this case, $R_+$ droplets expand in such a way that they finally invade
the system.

Hence, on generalizing this argument to arbitrary valued of $v$, one
obtains imposing $c_f=c$ in Eq. (\ref{cfront}) that the transition from 
convective to absolute instability occurs at:
\begin{eqnarray}
\epsilon_a &=& {3\over 16}(c^2 + {2\over \sqrt 3}vc -v^2) \, \,({\rm for}\,
 c < \sqrt 3 v)\nonumber \\
&=& {1\over 4}c^2 \, \,({\rm for}\,
 c > \sqrt 3 v )
\end{eqnarray}
From this result, it appears clearly that, when group velocity effects
dominate over nonlinear ones ($c > \sqrt 3 v$), the absolute instability
threshold remains the linear one. However, when nonlinearities dominate
($v>{c \over \sqrt 3 }$), the absolute instability threshold decreases,
but remains in the $\epsilon >0$ domain, when $v <c \sqrt 3$. It only
becomes negative when $v > c \sqrt 3$. This last case is the one
originally considered in \cite{chomaz}.

\subsection{Numerical Analysis}
\label{numeric0}

The above results have been checked through the numerical integration of the
equation (\ref{Rcglsub}). We will present here some of the data obtained for
systems being initially in the trivial steady state, and compare them to the
predictions obtained from the analytical analysis outlined in the preceeding
section. To observe a convective instability we consider a semi-infinite system
with one of the boundaries anchored to the unstable state $A(x=0)=0$. 
Experimentally, this boundary condition can be achieved using a negative value
for the control parameter $\epsilon$ for $x<0$.

The numerical integrations have been performed using a finite difference
method \cite{marc} with a spatial step of $\delta x=0.05$ and time step
$\delta t =0.001$, except where otherwise noted. As explained before, the 
boundary conditions for a system of size $L$ were taken as follows: 
$A=0$ at $x=0$ for all times and $\partial_x A=0$ at $x=L$. 

We only discuss here situations where the nonlinearities dominate over
mean flow effects, thus where linear instability criterion fails.

\begin{figure*}
\vspace*{-3.0cm}
\psfig{figure=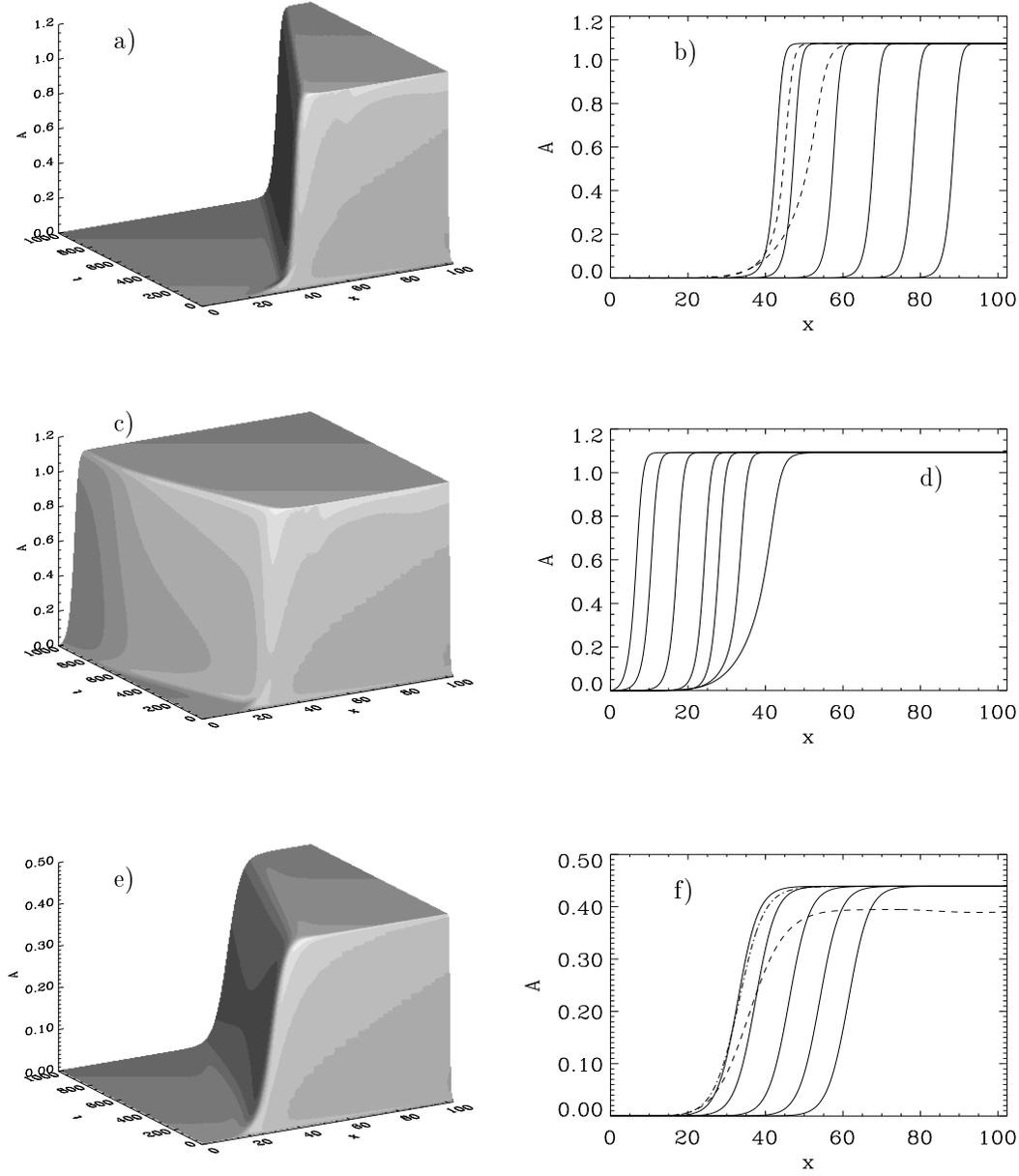,width=18cm}
\vspace*{-6.cm}
\caption{Deterministic evolution of perturbations of the trivial steady
state. The column on the left shows the spatio-temporal evolution of the
field $A$ and the column on the right the value of the field at different
times. The initial condition for each grid point is
$A_i(t=0)=f |\eta_i|$, where $\eta$ is a Gaussian random number of
zero mean and variance 1 and $f=10^{-4}$. In (a) and (b) we consider the
subcritical case in a convectively unstable regime, with $\epsilon=0.18$,
$v=1$, $c=1$. In (c) and (d) we consider the subcritical case in an
absolutely unstable regime, with $\epsilon=0.23$, $v=1$, $c=1$. In (e)
and (f) we consider the supercritical case, with $\epsilon=0.23$, $v=-1$,
$c=1$. Dashed lines in (b) and (f) correspond to early times when the
front is being formed and it effectively moves to the left. Continuous
lines show the front moving to the right in (b) and (f) and moving to the
left in (d)
\label{zerodeterministic}}
\end{figure*}

(1) A first case corresponds to $c=v=1$. In this case, the transition
from convective to absolute instability should occur at $\epsilon_a =
{\sqrt 3\over 8} \simeq 0.21$. This is illustrated by the numerical
results presented in fig. \ref{zerodeterministic}. In fig. 
\ref{zerodeterministic} (a) and (b), we show the deterministic evolution
of the field $A$ from random initial conditions around $A=0$ and for 
$\epsilon = 0.18$. The data confirm the convective nature of the 
instability. Effectively, we see, in a first stage, the building of a
front between the trivial state and the bifurcating one, and, in a second
stage, this front is advected out of the system. On the contrary, for
$\epsilon=0.23$, the instability is absolute, as shown in fig.
\ref{zerodeterministic} (c) and (d), where the front moves in the opposite
direction, and the bifurcating state invades the system. 

The difference between subcritical and supercritical behavior is
enlightened in fig. \ref{zerodeterministic} (e) and (f), where the field
evolution has been computed with the same parameters as in fig. 
\ref{zerodeterministic} (c) and (d), except that $v$ has been changed 
from $+1$ to $-1$ to simulate supercriticality. In this case, the 
instability should be convective, since the absolute threshold is 
$\epsilon = 0.25$, and the results are in agreement with this prediction.

\begin{figure}
\vspace*{-3.0cm}
\hspace*{-1.cm}
\psfig{figure=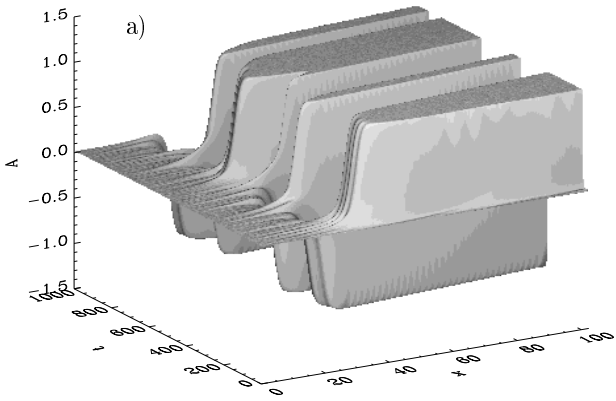,width=17cm}

\vspace*{-17.cm}
\hspace*{-1.cm}
\psfig{figure=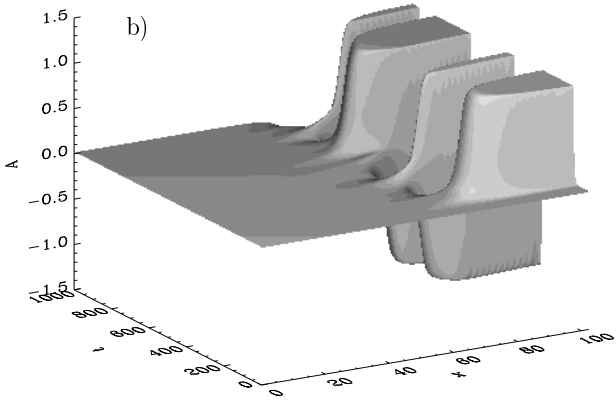,width=17cm}
\vspace*{-15.5cm}
\caption{Spatio-temporal evolution of the field $A$ with noise in a
convectively unstable regime for $\epsilon=0.18$, $v=1$, $c=1$. The initial
condition is $A(x,t=0)=0$. (a) noise intensity $\xi=10^{-6}$, (b) noise 
intensity $\xi=10^{-14}$.
\label{zerowithnoise}}
\end{figure}

The effect of noise in the regime of convective instability is presented
in fig. \ref{zerowithnoise}. The field dynamics has been computed for the
same values of the parameters as in fig. \ref{zerodeterministic} (a), but
in the presence of noise of different intensities. The noise intensity
has been fixed at $\xi=10^{-6}$ in fig. \ref{zerowithnoise} (a) and at 
$\xi=10^{-14}$ in fig. \ref{zerowithnoise} (b). In both cases we observe
noise sustained structures: Noise is able to sustain finite field
amplitudes (positive or negative, according to the $+,-$ symmetry of the 
system). Weaker noise induces larger healing length for the pattern.
Hence, in the stochastic case, pattern formation is sensitive to system
size, since the latter has to be larger than the healing length, for the
pattern to be able to develop.

\begin{figure}
\psfig{figure=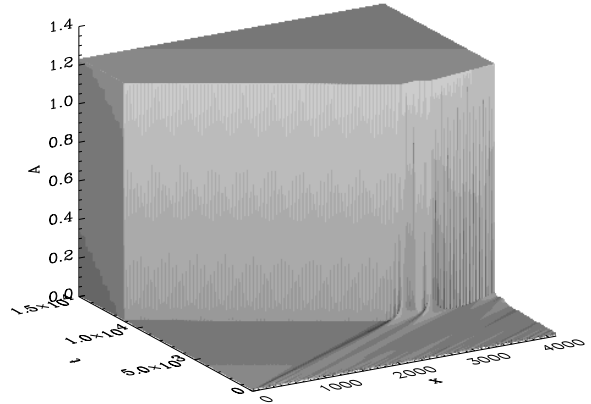,width=9cm}
\caption{Spatio-temporal evolution of the field $A$ without noise in an 
absolutely unstable regime with $\epsilon=0$, $v=1.5$, $c=0.5$.
The initial condition is as in Fig. \ref{zerodeterministic} but 
with $f=10^{-2}$. In this case we have taken $\delta x=1$, $\delta t=0.2$.
\label{epsilon=0}}
\end{figure}

(2) In a second case, we chose $c=0.5$, $v=1.5$, and this corresponds to 
the situation presented by Chomaz \cite{chomaz}, where nonlinear effects
dominate ($v > c \sqrt 3$.) and where the transition from convective to
absolute instability occurs in the subcritical domain since $\epsilon_a
\simeq -0.21$. For $\epsilon_a < \epsilon$, the instability is absolute,
but the dynamics is qualitatively different if $\epsilon$ is positive or
negative. When $\epsilon >0$, both linear and cubic terms are
destabilizing, and the building and propagation of fronts between trivial
and bifurcating states is much faster than for $\epsilon < 0$, when the
linear term is stabilizing, and the cubic one is destabilizing. When the
dynamics becomes very slow the time and system size needed to see the 
formation of a front from an initial perturbation become very large, so
that even in the absolutely unstable regime one might not observe the
decay of the state $A=0$ in finite times for a finite system. This effect
is illustrated in fig. \ref{epsilon=0} which corresponds to the absolutely
unstable regime. Note the significant increase of the times scales in 
comparison with figs. \ref{zerodeterministic} and \ref{zerowithnoise},
despite the fact that the perturbation of the zero state at the initial
time is much large (see the figure caption). We note that for $\epsilon
<0$ the evolution would still be slower. We first observe the formation
of the front ( initially moving to the right) and much later, when it
reaches the upper branch, invading the whole syste. Hence, when the
characteristic length needed for the building of the front is larger than
the system size, the instability is effectively convective, although the
system should be in the absolute instability regime (in the sense of
semi-infinite geometries). For the parameters chosen in this example and
for a length $L<2000$ one does not observe the decay of the state $A=0$.

It is worthnoting that the observed finite size effects confirm and
complement the analysis made by Chomaz and Couairon \cite{againstthewind} of
fully nonlinear solutions of Ginzburg-Landau equations in finite domains.
In case (1), for $\epsilon = 0.23$, nonlinear global (NLG) modes exist,
even in finite domains. However, since the basic state, $A=0$ is linearly
absolutely stable, NLG modes only develop if the initial condition is
sufficiently large for the transients to reach an order one amplitude in
the finite domain. Since the amplification factor increases exponentially
with $L$, the minimum amplitude of initial perturbations able to trigger
the NLG mode decreases exponentially with $L$ \cite{againstthewind}. As a
result, the development of NLG modes is almost insensitive, in most
practical situations, to system size.

On the contrary, in case (2), the basic state is linearly stable,
absolutely and convectively, and the minimum amplitude of initial
perturbations able to trigger NLG modes in finite boxes decreases linearly
with $L$. It is why, in the conditions of our numerical analysis, no global
mode is obtained for $L<2000$.

\section{Stability Analysis of the Bifurcating States}
\label{bifurcatingstate}

\subsection{Analytical Results}

The linear evolution around the upper branch steady states $R_+$ and
middle branch steady states $R_-$ is given by eq. (\ref{linampl}). The upper 
states $R_+$ are linearly stable for all $\epsilon > -{ v^2 \over 4}$.

On the other hand, the usual linear instability criterion shows that the
$R_-$ steady states are convectively unstable for $8R^2_-\sqrt { v^2 +
4\epsilon}< c^2$, and absolutely unstable for $8R^2_-\sqrt { v^2 +
4\epsilon} > c^2$. In other words, these states are absolutely unstable
in the range
\begin{equation}
-{1\over 8}(v^2 + {c^2\over 2 } + v \sqrt { v^2 -c^2})<\epsilon <
-{1\over 8}(v^2 + {c^2\over 2 } - v \sqrt { v^2 -c^2})
\end{equation}
and convectively unstable in the windows defined by
\begin{equation}
-{v^2\over 4 } <\epsilon < -{1\over 8}(v^2 + {c^2\over 2 } + v 
\sqrt { v^2 -c^2})
\end{equation}
and 
\begin{equation}
-{1\over 8}(v^2 + {c^2\over 2 } - v \sqrt { v^2 -c^2}) <\epsilon < 0
\end{equation}

\begin{figure}
\epsfig{figure=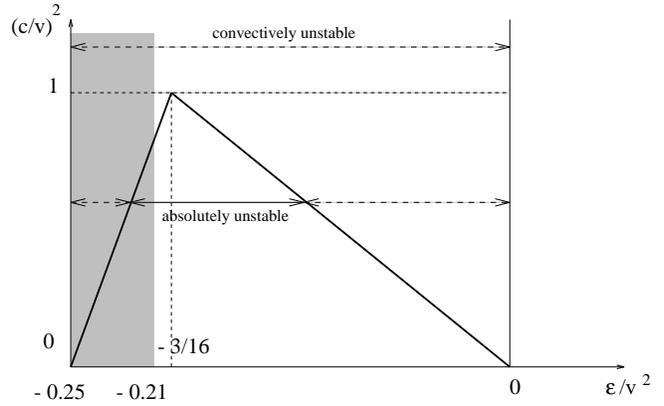,width=8.5cm}
\caption{Domains of linear convective and absolute instability of the
middle branch uniform steady state in the ($(c/v)^2$, $\epsilon/v^2$)
plane. The linear stability analysis is not valid in the hatched domain
where the nonlinearities are destabilizing.
\label{middlebranch}}
\end{figure}

Hence, when $v^2 < c^2$, these steady states are always linearly
convectively unstable. Still, when $v^2 > c^2$, there is a range of linear
absolute instability in the middle of their domain of existence, and a
range of linear convective instability close to the points where these
states disappear. This is shown in fig. \ref{middlebranch}.

Nevertheless the linear stability criteria may fail in the presence of
destabilizing nonlinearities. This is not only the case for the evolution of
the perturbations of the trivial steady state since the bifurcation is
subcritical, but it may also be the case for the perturbations around the
middle steady state branch, whose evolution is given by
\begin{eqnarray}\label{lowerstatev}
\partial_t \rho_- &+& c\partial_x \rho_- = + 2R^2_-\sqrt { v^2 + 4\epsilon}
\,\rho_- + \partial^2_x\rho_-\nonumber\\
&-& R_-(2v-5\sqrt { v^2 + 4\epsilon})\,\rho_-^2 -
(4v-5\sqrt { v^2 + 4\epsilon})\,\rho_-^3 \nonumber\\
&-& 5R_-\rho_-^4 - \rho_-^5
\end{eqnarray}
The quadratic nonlinearity is destabilizing for $ \epsilon <
\epsilon_L = -0.21\, v^2$. In such cases, one has to perform a nonlinear
analysis of the dynamics to determine the convective or absolute nature of
the instability.

In the regime where the nonlinearities of the evolution equation
(\ref{lowerstatev}) are stabilizing, i.e. for $ \epsilon_L = -0.21\, v^2
<\epsilon < 0 $, the results of the linear analysis may be assumed to be
valid. Hence, we may safely rely on these results above the metastability
point, i.e. for $ \epsilon_M = -3/16 v^2 <\epsilon < 0 $. Below the
metastability point, i.e. for $ -0.25 v^2 <\epsilon < -3/16 v^2 $, one
has to perform a nonlinear analysis, which, in this case, relies on the
evolution of fronts between middle branch states and the trivial steady
state. We do not perform this analysis here since it would only affect 
quantitatively but not qualitatively the results presented above.

\subsection{Dynamics of the Subcritical Unstable Branch}

We have numerically confirmed the convective nature of the instability of the
subcritical middle branch. The numerical integration has been performed as
indicated in subsection \ref{numeric0}. Also, as indicated in that subsection,
to observe a convective instability we consider a semi-infinite system with
one of the boundaries anchored to the unstable state. Here we have to take
$A(x=0)=R_-$ corresponding to the field amplitude of the subcritical middle
branch. Experimentally, this boundary condition can not be achieved as easily
as before because there is no value of the control parameter $\epsilon$ for
which $A(x)=R_-$ is an homogeneous steady stable state. However depending on
the system it can be imposed in different ways. In an optical system, for
example, the left boundary condition could be achieved injecting an external
field at $x=0$ with the apropriate amplitude. Finally, as in subsection
\ref{numeric0}, the right boundary condition is taken as $\partial_x A=0$ at
$x=L$

We computed the evolution from an initial steady state with $R^2_- = 0.1$
on the middle branch, which corresponds to $v=1$ and $\epsilon = -0.09$.
We then study the system dynamics for different values of the group
velocity $c$. According to the previous discussion, for

(1) $c< 0.8$, the state $R_- $ should be absolutely unstable

(2) $c> 0.8$, the state $R_- $ should be convectively unstable.

\begin{figure}
\vspace*{-3.0cm}
\hspace*{-1.cm}
\psfig{figure=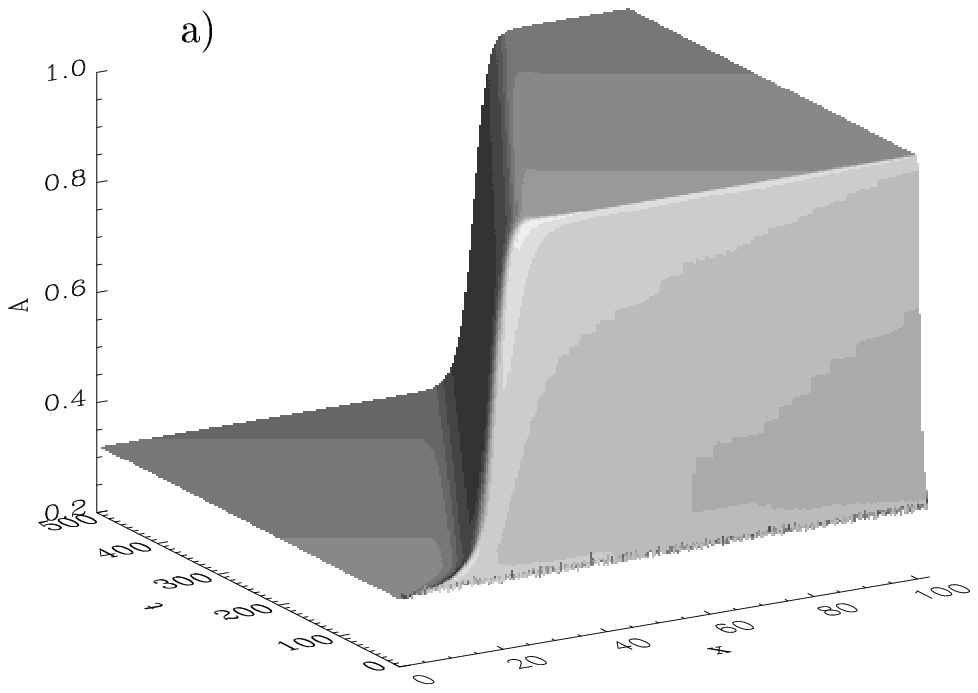,width=17cm}

\vspace*{-17.cm}
\hspace*{-1.cm}
\psfig{figure=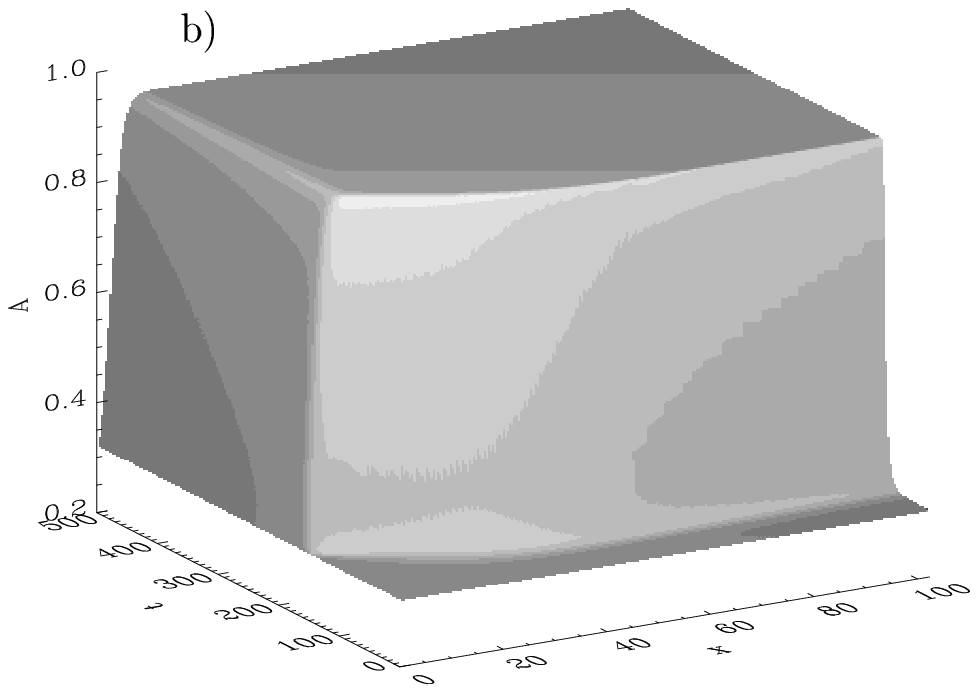,width=17cm}
\vspace*{-15.5cm}
\caption{Deterministic evolution of a perturbation around the middle
branch state $R_-$ in a convectively unstable regime for
$\epsilon=-0.09$, $v=1$. The initial condition for each grid point is
$A_i(t=0)= R_- + f|\eta_i|$, where $\eta$ is a Gaussian random number
of zero mean and variance 1 and $R_-=\sqrt{0.1}$. (a) convectively
unstable regime with $c=1$ and $f=10^{-2}$. (b) absolutely unstable regime
with $c=0.55$ and $f=10^{-6}$.
\label{middledeterministic}}
\end{figure}

In fig. \ref{middledeterministic} (a), we present, for $c=1$, the results
obtained for the deterministic evolution of an initial perturbation of
the state $R_-$. They show that the instability is effectively 
convective. On lowering the group velocity from $c=1$ to $c=0.55$, the 
nature of the instability changes from convective to absolute, as
expected, and shown in fig. \ref{middledeterministic} (b). These results
confirm that the middle branch, which is always stable for $c=0$, may be
stabilized by mean flow effects in deterministic systems, in the sense
that there is a range of parameters in which it is only convectively 
unstable. 

\begin{figure}
\psfig{figure=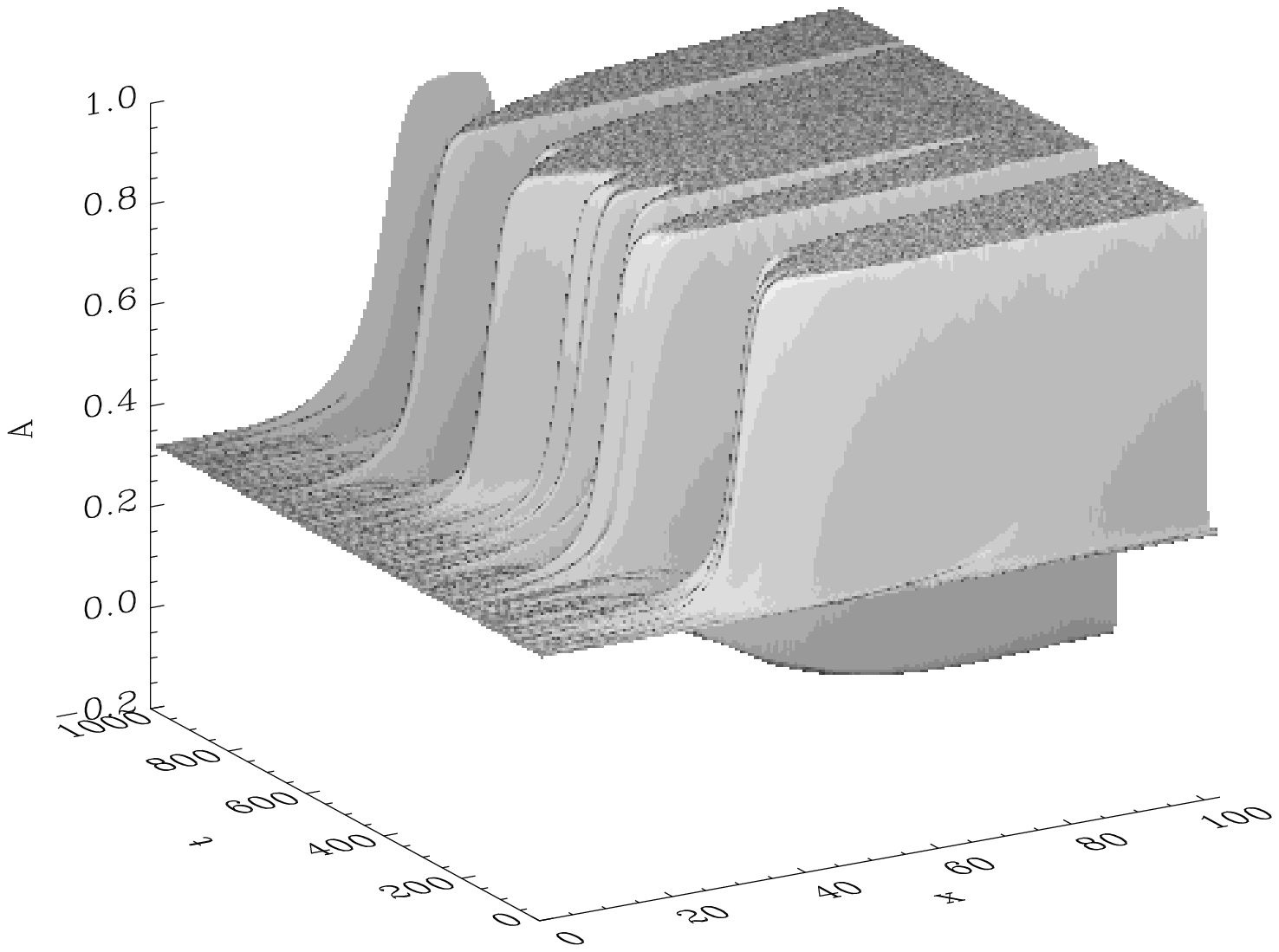,width=9cm}
\caption{Spatio-temporal evolution of the field $A$ with noise in a 
convectively unstable regime for the middle branch for $\epsilon=-0.09$, 
$v=1$, $c=1$ and $\xi=10^{-6}$. The initial condition is $A(x,0)=
\sqrt{0.1}$.
\label{middlenoise}}
\end{figure}

The effect of noise in the convectively unstable regime of the trivial
state was to sustain a structure continuously excited by noise. In the
case of the middle branch. $R_-$, and when this is convectively unstable,
noise forces the system to relax randomly to either of the two coexisting
stable branches, as shown in fig. \ref{middlenoise}. Still, if noise is
weak in comparison with the strength needed to see its effect in a finite
system, one would observe the middle branch as effectively stable.

\section{Conclusions}
\label{conclusions}

In this paper, we considered systems described by the subcritical 
Ginzburg-Landau equation, and analyzed some problems related with the
effect of group velocities on the stability of its steady states. In the
case of the trivial steady state, it is known that the transition between
convective and absolute linear instability regimes is shifted by the
effect of destabilizing nonlinearities, and the corresponding nonlinear
absolute instability threshold may easily be computed for semi-infinite
systems \cite{chomaz,couairon}. Our numerical study of the evolution of
perturbations from the trivial steady state in finite systems shows that,
in a first step, a front is built between this state and the bifurcating
one, which corresponds to the upper branch of steady states. Then,
according to the intensity of the group velocity, the front moves outwards
or inwards, which corresponds to convective or absolute instability,
respectively. When the characteristic length needed for the building of
the front is shorter than the system size, the nature of the instability
is in agreement with the theoretical predictions made for semi-infinite
systems. However, our numerical results show that, if the characteristic
building length of the front is larger than the system size, one will
never see inward motion of the front, and, in this case, even above the
absolute instability threshold, the instability is effectively
convective.

We also studied the instability of the subcritical middle branch of steady
states, a problem that had not been addressed up to now. It may be shown,
already at the level of a linear analysis, that this branch, which is
absolutely unstable without group velocity, may entirely become
convectively unstable in the presence of group velocities larger than
some well-defined critical value. This result has been confirmed by the
numerical analysis of the evolution of perturbations of steady states on
this branch. The stabilization of such steady states has effectively be
obtained, in deterministic systems, for group velocities in the
predicted range. In stochastic systems, however, these steady states relax
to one of the stable branches, as expected. Nevertheless, for this
relaxation to occur, either noise strength or system size have to be large
enough. This effect may be of practical importance, for example, in binary
fluid convection, where, besides the fact that the role of subcriticality
is not clearly understood yet \cite{kolodner}, the presence of natural or
forced mean flows, or group velocities, could effectively stabilize
otherwise unstable branches of steady states.

\section{Acknowledgements}
Financial support from DGICYT (Spain) Project PB94-1167 is acknowledged.
DW is supported by the Belgian National Fund for Scientific Research.
The authors also acknowledge helpful discussions with W. van Saarloos.

\pagebreak

\end{document}